\documentstyle[12pt]{article}

\def\be{\begin{equation}}
\def\ee{\end{equation}}
\def\ltap{\raisebox{-.55ex}{\rlap{$\sim$}} \raisebox{.4ex}{$<$}}
\def\gtap{\raisebox{-.55ex}{\rlap{$\sim$}} \raisebox{.4ex}{$>$}}
\def\gsim{\mathrel{\gtap}}
\def\lsim{\mathrel{\ltap}}
\def\Tr{\mbox{Tr}}

\begin{document}
\begin{flushright}
PURD-TH-95-01 \\
April 1995  \\
hep-ph/9504389
\end{flushright}
\vspace{0.4in}
\begin{center}
{\Large
Strong-coupling expansions for chiral models of electroweak symmetry
breaking} \\
\vspace{0.3in}
S.Yu. Khlebnikov\footnote{
Alfred P. Sloan Foundation Fellow; DOE Outstanding Junior Investigator}
and R.G. Schnathorst
\\
\vspace{0.1in}
{\it Department of Physics, Purdue University,
West Lafayette, IN 47907, USA} \\
\vspace{0.4in}
{\bf Abstract}
\end{center}
We consider chiral $U(N)\times U(N)$ models with fermions in the limit of
infinitely large local bare Yukawa coupling. When the scalar field is subject
to non-linear constraint, phase transitions in these models are seen to be
identical to those in the corresponding purely bosonic ones.
Relaxing the non-linear constraint, we compute the seventh-order
strong-coupling series for the susceptibility in these models and analyze
them numerically for the $U(2)\times U(2)$ case. We find that in four
dimensions the approach to the phase transition follows to a good accuracy
the mean-field critical behavior, indicating the absence of non-trivial
fixed points at strong coupling and being consistent with the first-order
nature of the transition. In three dimensions, the strongly-coupled
bosonic $U(2)\times U(2)$ model (without gauge fields) has a first-order
transition strong enough to accommodate electroweak baryogenesis only for
a narrow region of the bare parameter space. \\
\newpage
Phase transitions in chiral sigma models are of interest both
in three dimensions where they serve as models of the chiral phase
transition in QCD \cite{Wilczek}, and in four dimensions where the
nature of the phase transition
determines viability of certain models of dynamical electroweak symmetry
breaking \cite{CGS,Shen,BHJ,CL}.
In the latter case, the analysis should include fermions with large Yukawa
couplings because they
can influence the nature of the transition. In ref.\cite{BHJ}, it was argued,
based on an approximate method of dealing with strongly coupled fermions,
that a large Yukawa interaction can increase an otherwise unacceptably
small hierarchy between the symmetry breaking scale and the cutoff.
Ref.\cite{CL} considered the case of moderate ($O(1)$) bare Yukawa coupling
and found no substantial increase in the hierarchy for that case.

In this note we consider $U(N)\times U(N)$ chiral models with fermions in
the limit of infinite bare Yukawa coupling. This limit should
give us some insight into the behavior of systems with finite but large
(much larger than one) bare Yukawa coupling $y$.
For models where the order parameter is a single scalar field, it is known that
the phase diagram in the limit $y\to \infty$ depends on the choice of a lattice
form for the Yukawa interaction \cite{LSS}. This is natural when there are
phase transitions that involve both ferromagnetic and antiferromagnetic order,
because in that case the order parameter varies on the scale of lattice spacing
and that variation cannot be removed by a redefinition of the field. Here we
are interested in a paramagnetic-ferromagnetic transition, so we should choose
a lattice action that has it.

We use the local form of Yukawa interaction
\be
S_Y = y \sum_i ({\bar \chi}^a_i \phi_i \psi^a_i +
{\bar \psi}^a_i \phi_i^{\dagger} \chi^a_i)
\label{y}
\ee
where $i$ labels sites of a four-dimensional lattice with unit spacing;
$\psi^a_i$ (${\bar \psi}^a_i$) and $\chi^a_i$ (${\bar \chi}^a_i$)
are staggered fermions, each has one spin
component per site but forms a fundamental (conjugated fundamental)
representation of the respective
$U(N)$ and in addition comes in $N_c$ "colors", $a=1,...,N_c$.
This corresponds to having $8N_c$ Dirac fermions in a continuum theory.
Just as in the case of a single scalar \cite{LSS}, in the limit $y\to\infty$
the fermion fields can be integrated away leaving a local correction to the
potential for $\phi$. In our case, this changes scalar
potential $V(\phi)$ into
\be
{\tilde V}(\phi) = V(\phi) - N_c \Tr\ln(\phi_i^{\dagger} \phi_i) \; .
\label{eff}
\ee
This result holds when fermions have a gauge interaction.

In the {\em non-linear} limit of the model, when the order parameter is
subject to the constraint $\phi_i^{\dagger} \phi_i\propto 1$, the correction
to the potential is a
constant. We see that bosonic correlators of a non-linear model
with $y\to\infty$ coincide with those of the
the corresponding model without fermions. Hence, the phase transitions in
the two models are identical. If the phase transition in the model without
fermions is of the first order, as suggested by the previous work, so is the
phase transition in the model with fermions in the limit of infinite
Yukawa coupling, and with exactly the same strength. (This applies
also to models with an hermitean order parameter considered in
ref.\cite{herm}.)

Even when the non-linear constraint is relaxed, there can be no second-order
or weakly first-order ferromagnetic-paramagnetic
transition in a $y\to\infty$ model if there is none
in its bosonic counterpart. This is a simple consequence of universality,
since the logarithmic interaction in (\ref{eff}) is a local one.
So, for the purpose of searching for non-trivial fixed points and weakly
first-order transitions, it is in principle sufficient to study
the purely bosonic model, without the logarithm, although we have also done
independent analysis of the effective theory with the potential ${\tilde V}$.

We have computed the strong-coupling expansion for
the effective bosonic theory to the seventh order and done numerical
analysis of the series for the $U(2)\times U(2)$ case. Our results are
consistent with the absence of non-trivial fixed points
at strong coupling both in the case with and
without fermions. Assuming that the phase transition is of the first order,
we find that it is a relatively weak one for most of the bare parameter
space. We also considered the $U(2)\times U(2)$ model without fermions in
three dimensions
and found that except for a narrow region of bare parameters near
the line beyond which the potential becomes unbounded from below, the phase
transition (without gauge fields) is too weakly first-order to
provide deviations from equilibrium required for electroweak
baryogenesis \cite{baryo}.

Thus, we consider the linear $U(N)\times U(N)$ model for which the bosonic
action $S$ is
\begin{eqnarray}
S & = & -\sum_{\langle ij \rangle} [\Tr(\phi_i^{\dagger} \phi_j) +
\mbox{h.c.}]  + \sum_i V(\phi_i)  \label{lin} \\
V(\phi_i) & = &  -\mu^2 \Tr(\phi_i^{\dagger} \phi_i) +
\lambda_1 [\Tr(\phi_i^{\dagger} \phi_i)]^2 +
\lambda_2 \Tr(\phi_i^{\dagger} \phi_i)^2 \label{pot} \; ;
\end{eqnarray}
$\phi$ is now an arbitrary complex $2\times 2$ matrix.
We consider only the case $\lambda_2 > 0$. The bare action (\ref{lin})
is then bounded from below if $N\lambda_1+\lambda_2>0$.
The non-linear limit is recovered if we let $\mu^2$, $\lambda_2$ and
$N\lambda_1+\lambda_2$ to infinity with $\mu^2/(N\lambda_1+\lambda_2)$ fixed.

The strong coupling expansion for (\ref{eff}) is the expansion in powers of
the first term in (\ref{lin}).\footnote{In statistical mecnahics, such
expansions are called high-temperature expansions \cite{exp}. We chose
not to use that terminology to avoid confusion with field theories at finite
temperature.}
Coefficients in that expansion are functions of the bare parameters
$\mu^2$, $\lambda_1$ and $\lambda_2$ and can be expressed through
linear combinations of products of invariant ordinary (non-functional)
integrals such as
\begin{eqnarray}
\nonumber \\
I_1 & = & Z^{-1} \int \Tr(\phi^{\dagger} \phi)
\exp[-{\tilde V(\phi)}] d\phi \; ,
\nonumber \\
I_2 & = & Z^{-1} \int \Tr(\phi^{\dagger} \phi)^2
\exp[-{\tilde V(\phi)}] d\phi \; ,
\label{inv} \\
I_3 & = & Z^{-1} \int [\Tr(\phi^{\dagger} \phi)]^2
\exp[-{\tilde V(\phi)}] d\phi \; ,
\nonumber \\
I_4 & = & Z^{-1} \int \Tr(\phi^{\dagger} \phi)^3
\exp[-{\tilde V(\phi)}] d\phi  \; ~~~\mbox{etc.}
\nonumber
\end{eqnarray}
where ${\tilde V}$ is given in (\ref{eff}) and
$Z = \int \exp[-{\tilde V(\phi)}] d\phi$.

The invariant integrals $I_n$ of (\ref{inv}), as functions of the bare
parameters, have the form
\be
I_n = (2 \lambda_s)^{-p/2} F_n(\lambda_2/\lambda_s, \mu^2/\sqrt{2 \lambda_s})
\label{form}
\ee
where
\[
\lambda_s=\lambda_1+\lambda_2
\]
and $p$ is the total power of
$\phi$ and $\phi^{\dagger}$ in the traces in the integrand: $p=1$ for $I_1$,
$p=2$ for $I_2$ and $I_3$, $p=3$ for $I_4$ etc. So, for fixed
$\lambda_2/\lambda_s$ and $\mu^2/\sqrt{2 \lambda_s}$, the strong coupling
series is a series in powers of $\beta=(2 \lambda_s)^{-1/2}$.

We have computed the seventh-order series for zero-momentum susceptibility
\be
\chi=\sum_i\langle \Tr\phi_0 \phi^{\dagger}_i \rangle
\label{sus}
\ee
for three- and four-dimensional lattices in terms of the invariant integrals
for a general $U(N)\times U(N)$ model with infinite bare Yukawa coupling.
Calculation of the integrals and analysis of the series were done
for the $U(2)\times U(2)$ case. The series were obtained by the recursion
method of ref.\cite{SK}. The complete series will be presented elsewhere;
here we limit ourselves to the results of the analysis.\footnote{
Comparison of our series to known limiting cases has found
one discrepancy. The $N=2$ case with $\lambda_2=0$ includes the $O(8)$ model.
For that model on the fcc (face-centered cubic) lattice we obtain the 7th
order coefficient of 14492289.1770 while Table 8.5 of ref.\cite{Stanley} lists
14490203.7349. Other coefficients through the 7th order for that model
on the fcc and bcc (body-centered cubic) lattices agree. Our computer program
generating the series has also reproduced, through the 7th order, the
known results for models with fewer components of order parameter:
the classical $XY$ model \cite{BJ} and the general one-component model
\cite{scalar} on various three-dimensional lattices, and the Ising model
on simple and face-centered hypercubic four-dimensional lattices \cite{ising}.
This gives us confidence that our program generates series correctly.}

Let us consider first the purely bosonic model ($N_c=0$) on
the four-dimensional simple hypercubic lattice. The values of bare parameters
for which Shen \cite{Shen} finds evidence for a first-order
transition at large bare couplings correspond in our notation\footnote{
Our $\mu^2$ is $-m^2$ of ref.\cite{Shen} minus the number of nearest neighbors;
our $\lambda_1$ and $\lambda_2$ are four times those of ref.\cite{Shen}.}
to $\lambda_2/\lambda_s=1$ ($\lambda_1=0$)
and $\mu^2/\sqrt{2 \lambda_s}=2.41775$.
With these parameters, the series for the susceptibility for
the $U(2)\times U(2)$ model is
\begin{eqnarray}
\chi/\chi_0 & = & 1 + 9.98596 \beta + 90.66614 \beta^2 + 822.72732 \beta^3
+ 7347.40485 \beta^4
\nonumber \\
 & & \mbox{} + 65668.03299 \beta^5 + 583436.92981 \beta^6
+ 5186731.98528 \beta^7 + ...
\label{ser}
\end{eqnarray}
where $\beta=(2\lambda_s)^{-1/2}$ and $\chi_0$ is the zeroth-order
susceptibility which has been factored out.

The series were analyzed using Zinn-Justin's method \cite{ZJ,Guttmann}.
Denote the coefficients in (\ref{ser}) as $a_n$, so that
\be
\chi/\chi_0 = \sum_{n=0}^{\infty} a_n \beta^n \; .
\label{den}
\ee
In Zinn-Justin's method one forms the ratios
\be
s_n=-\left( \ln\frac{a_n a_{n-2}}{a_{n-1}^2} \right)^{-1} \; .
\label{rat}
\ee
Then, estimates of the susceptibility exponent $\gamma$ are obtained as
\be
\gamma_n = 1+2\frac{s_n+s_{n-1}}{(s_n-s_{n-1})^2} \; ,
\label{gam}
\ee
and estimates of the critical "temperature" as
\be
\beta_{c,n}^{-1} = \left( \frac{a_n}{a_{n-2}} \right)^{1/2}
\exp\left[-\frac{{s_n+s_{n-1}}}{(s_n-s_{n-1}) s_n} \right] \; .
\label{tem}
\ee
For the series (\ref{ser}) we obtain
\begin{eqnarray}
\{\gamma_3,...,\gamma_7\} & = &
\{1.0011, 1.0013, 0.9986, 0.9989, 0.9991\} \; , \label{estg} \\
\{\beta_{c,3}^{-1},...\beta_{c,7}^{-1}\} & = &
\{9.072, 9.158, 8.940, 8.952, 8.892\} \; . \label{estt}
\end{eqnarray}
We observe that the estimates for $\gamma$ are very close to 1 and are
remarkably stable for such a short series.

The values of couplings $\lambda_1$ and $\lambda_2$ to which a
given estimate refers are determined a posteriori, through the critical
value $\beta_c$.
Our estimates for $\beta_c$ and the critical exponent become less stable
as we try to move in the region of larger $\beta_c$ or, equivalently,
smaller $\lambda_s$. As a result, we could not probe the weak
coupling region $\lambda_s\lsim 1$. For the purely bosonic
theory, the usual perturbation is applicable for small $\lambda$, so this
is not a very significant limitation. In the strong-coupling region
$\lambda_s\gsim 1$, except for a narrow region near the line
$2\lambda_1+\lambda_2=0$, stable estimates were obtained and they were
consistent with $\gamma=1$. We interpret these results as follows.

At a second-order phase transition, the correlation length grows to infinity
and susceptibility is singular, $\chi\sim (\beta_c-\beta)^{-\gamma}$.
However, even for a first-order transition, as it is approached,
the correlation length and the susceptibility may grow
somewhat before the transition takes place. $\gamma=1$ is the mean-field
value and it implies, in the renormalization group language, that in our
case the approach to the phase transition is controlled by the trivial fixed
point $\lambda_1=\lambda_2=0$. This fixed point is known to be infrared
unstable by perturbation theory \cite{Wilczek} --- the signal of a first-order
transition.

A seventh-order series probes distances of about seven lattice
spacings. If there were an infrared fixed point with $\gamma\neq 1$ at
strong coupling, we could detect it even with such a short series by taking
the bare couplings close to the fixed point.
Because non-trivial critical behavior was not found
for {\em any} $\lambda_1$, $\lambda_2$ with large $\lambda_s$,
we can claim the absence of strongly-coupled infrared fixed points,
except for the unlikely case that such a fixed point has $\gamma$ very
close to 1.

If we assume that the phase transition is, in fact, of the first order,
then for the $U(2)\times U(2)$ model it is a relatively weak one for most
of the bare parameter space. Indeed, we have seen
that the susceptibility follows the mean-field, second-order, behavior up
to rather large scales. Equivalently, the correlation length grows
to the size of at least several lattice spacings.
It is also instructive to compare the estimates (\ref{estt}) for
the critical "temperature" $\sqrt{2\lambda_s}$ of what our series
sees as a second-order transition with the value for which ref.\cite{Shen}
finds a first-order transition. In our normalization, that value is
$\sqrt{2\lambda^{f.o.}_s}=2\sqrt{20}=8.94427$, rather close to the numbers
in (\ref{estt}). The fact that mistaking
the first-order transition for a second-order one makes only a small error
in the critical "temperature" confirms that the transition is only weakly
first-order.

The correlation length for electroweak interactions is the scale of physics
responsible for the symmetry breaking and cannot be much above, say, 2 TeV.
Therefore, an hierarchy between the correlation length and the cutoff
by a factor of 10 or so should be sufficient for consistency of models
of electroweak symmetry breaking in which the cutoff scale is not much
above 20 TeV.

Near the line $2\lambda_1+\lambda_2=0$, beyond which the potential becomes
unbounded from below, the transition becomes more strongly first-order.
We will describe the change in the behavior of the series in that
region while discussing the three-dimensional case, for which this
change may have an application to electroweak baryogenesis.

In three dimensions, our results indicate that for
small $\lambda_2$ and large $\lambda_1$, the approach to the phase
transition, at the length scale probed by the 7th order series, is
controlled by the non-trivial $O(8)$ fixed point, but for most of
the bare couplings plane, by the trivial point $\lambda_1=\lambda_2=0$.
Assumimg that the transition is of the first order, we find
that it remains a weak one, except for the vicinity of
$2\lambda_1+\lambda_2=0$. For the simple cubic lattice and
$\lambda_2/\lambda_s=0.5$, $\mu^2/\sqrt{2 \lambda_s}=-5.$, corresponding to
the values of fig.3 of ref.\cite{Shen3}, we obtain
\begin{eqnarray}
\{\gamma_3,...,\gamma_7\} & = &
\{1.0005, 1.0005, 1.0009, 1.0007, 0.9997\} \; , \label{estg3} \\
\{\beta_{c,3}^{-1},...\beta_{c,7}^{-1}\} & = &
\{1.023, 1.030, 1.011, 1.008, 1.020\}   \label{estt3}
\end{eqnarray}
in good agreement with the mean field value $\gamma=1$ and the "temperature"
of the first order transition $\sqrt{2\lambda^{f.o.}_s}\approx 1$
deduced from fig.3 of ref.\cite{Shen3}.
Similar results were obtained for stronger couplings. This is in accord
with the results of a numerical simulation of the
non-linear $U(2)\times U(2)$ model in three dimensions \cite{Dreher}, which
has found that to observe the first-order phase transition one has to go to
lattices as large as $16^3$.

Near the line $2\lambda_1+\lambda_2=0$, or $\lambda_2/\lambda_s=2$,
the behavior of the series changes. Compare the following estimates for
the critical exponent obtained for the simple cubic lattice,
$\mu^2/\sqrt{2 \lambda_s}=-2.$ and different values of $\lambda_2/\lambda_s$:
for $\lambda_2/\lambda_s=1.55, 1.6, 1.65$, respectively,
\begin{eqnarray*}
\{\gamma_3,...,\gamma_7\} & = &
\{0.9996, 0.9996, 0.9990, 0.9992, 0.9991 \}  \; , \\
\{\gamma_3,...,\gamma_7\} & = &
\{0.9992, 0.9993, 0.9988, 0.998 + 1.45\times 10^{-7} i,
3.16 + 0.886 i \} \; , \\
\{\gamma_3,...,\gamma_7\} & = &
\{2.91+0.218 i, 5.09, -1.47+1.34 i, -31.5-533. i, 8.38+2.55 i \} \; .
\end{eqnarray*}
Imaginary values reflect the appearance of negative coefficients in the series
which starts to display irregular behavior; the prescription for
determining the signs of imaginary parts was chosen arbitrarily.
We see that as we approach the
line $\lambda_2/\lambda_s=2$, the irregular behavior of the series sets off in
lower order. In some systems, the presence of negative coefficients
in a series signals an unphysical singularity close to $\beta=0$, which can be
mapped further away by an appropriate change of the expansion variable
\cite{Guttmann}. In our case, there is a physical reason for such behavior ---
the decrease in the correlation length as the phase transition becomes more
strongly first-order.

For the electroweak phase transition, the cutoff in the effective
three-dimensional theory is of order of temperature. To prevent the washout
of baryon asymmetry, the expectation value of the order parameter after the
transition should be at least of order of temperature \cite{baryo} and,
at strong coupling, so should be the inverse correlation lengths in both
phases. So, we find that for the strongly-coupled $U(2)\times U(2)$ model
without gauge fields, the condition for electroweak baryogenesis is realized
only in a narrow region near the line $2\lambda_1+\lambda_2=0$.

The phase transition becomes more strongly first-order for larger
$U(N)\times U(N)$ groups \cite{Dreher,ASS}. However, from the results
of ref.\cite{Dreher} we were unable to conclude whether for $N=3$ it is
strong enough to preserve the baryon asymmetry for general values of the
couplings.

Finally, let us turn to the $y\to\infty$ limit of the four-dimensional
$U(2)\times U(2)$ model with fermions.
Though universality implies that there can be no infrared fixed points or weak
first-order transitions other than those of the purely bosonic model,
we have done an independent series analysis.
The results are indeed similar to those in the bosonic case.
Stable estimates of $\gamma$ were obtained for different $N_c$
for most of the strong-coupling region, $\lambda_s\gsim 1$,
and they were consistent with $\gamma=1$. The phase transition becomes
more strongly first-order in the vicinity of the line $2\lambda_1+\lambda_2=0$.
Similar results were obtained for the face-centered hypercubic lattice
(with naive fermions), although Zinn-Justin's method often did not lead to
stable estimates of $\gamma$ in that case.

To summarize, expanding in the inverse bare Yukawa constant $y$ gives
the effective bosonic theory (\ref{eff}) for the $y\to\infty$ limit
of chiral models of electroweak symmetry breaking that contain heavy fermions.
The non-linear limit of that effective theory coincides with that of
the model without fermions. For the linear $U(2)\times U(2)$ model in four
dimensions, the analysis of the seventh-order strong coupling series for
the susceptibility is consistent with the absence of any non-trivial fixed
points at large bare self-coupling, both in the case without fermions and
in the effective theory of the $y\to\infty$ limit.
Assuming that the phase transition in the strongly-coupled
$U(2)\times U(2)$ model is of the first order, it is a relatively
weak one, both in four and (in the case without fermions) three dimensions,
except for a region near the line beyond which the bare potential is unstable.

We are grateful to T. Clark and S. Love for getting us interested in the
subject and many helpful discussions. S.K. was supported in part by
Alfred P. Sloan Foundation and in part by the U.S. Department of Energy under
grant DE-AC02-76ER01428 (Task B). R.S. was supported in part by a grant
from Purdue Research Foundation.

\end{document}